\documentclass[twocolumn,secnumarabic,amssymb,showpacs,showkeys, nobibnotes, aps, prd]{revtex4}

\usepackage[dvips]{graphicx}
\newcommand{\be}{\begin{equation}}
\newcommand{\ee}{\end{equation}}
\newcommand{\bea}{\begin{eqnarray}}
\newcommand{\eea}{\end{eqnarray}}
\newcommand{\bm}{\begin{mathletters}}
\newcommand{\eml}{\end{mathletters}}

\newcommand{\oh}{\frac{1}{2}}
\newcommand{\nn}{\\ \nonumber}

\newcommand{\ov}{\overline}

\newcommand{\la}{\langle}
\newcommand{\ra}{\rangle}

\begin{document}

\title{Systematics of $2^+_1$ states in semi-magic nuclei}

\author{D.S. Delion$^{1,2}$ and N.V. Zamfir$^{1}$}

\affiliation{
$^1$ "Horia Hulubei" National Institute of Physics and Nuclear Engineering, \\
407 Atomi\c stilor, Bucharest-M\u agurele, 077125, Rom\^ania \\
$^2$ Academy of Romanian Scientists, 54 Splaiul Independen\c tei,
Bucharest, 050094, Rom\^ania}

\begin{abstract}
We propose a simple systematics of low lying $2^+_1$ energy levels and 
electromagnetic transitions in semi-magic isotopic chains $Z=28,50,82$ 
and isotonic chains $N=28,50,82,126$. To this purpose we use a two-level 
pairing plus quadrupole Hamiltonian, within the spherical 
Quasiparticle Random Phase Approximation (QRPA).
We derive a simple relation connecting the $2^+_1$ energy with the 
pairing gap and quadrupole-quadupole (QQ) interaction strength.
It turns out that the systematics of energy levels and B(E2) values
predicted by this simple model is fulfilled with a reasonable accuracy 
by all available experimental data. Both systematics suggest 
that not only active nucleons but also those filling closed shells
play an important role.
\end{abstract}

\pacs{21.10.Re, 21.60.Jz, 23.20.Lv}

\keywords{energy levels, electromagnetic transitions, semi-magic 
nuclei, Quasiparticle Random Phase Approximation}

\maketitle

Semi-magic nuclei are important for nuclear structure studies, 
due to the fact that relative simple shell-model 
configurations are used in describing collective states. 
The $2^+_1$ energy in semi-magic nuclei within
the generalized seniority scheme is constant along a 
given isotopic/isotonic chain \cite{Tal71}. 
Talmi's classical example is Sn isotopic chain
where the $2^+_1$ energy is fairly constant throughout the complete
shell from N=52 up to N=80 \cite{Cas00}. However, the $2^+_1$ energy
in other semi-magic nuclei is not a constant and depends on the
matrix elements of the effective interaction.

There are several systematic studies concerning spectroscopic
properties of even-even nuclei 
\cite{Ant68,Yam68,San97,Ter02,Sar08,Del10}.
In this paper we will investigate low lying $2^+_1$ states in
semi-magic isotopic and isotonic chains, 
by using a spherically symmetric Hamiltonian 
with two levels, one for protons and one for neutrons. 
In our calculations we have used the experimental binding energies
\cite{Mol95} in order to estimate the pairing gaps, which have 
nonvanishing values for magic proton or neutron numbers.
Thus, we include both kinds of nucleons because not only 
active particles but also the  particles filling the closed shell are 
important in describing low lying collective states.
The Hamiltonian contains the most relevant degrees of freedom,
given by the pairing and quadrupole-quadrupole (QQ) two body 
interactions. By using the standard quasiparticle representation one has
\bea
H&=&E_pN_p+E_nN_n-
\oh(
\kappa_pQ_{p}Q^{\dag}_{p}+\kappa_nQ_{n}Q^{\dag}_{n}
\nn&+&
\kappa_{pn}Q_{p}Q^{\dag}_{n}+\kappa_{pn}Q_{n}Q^{\dag}_{p})~,
\eea
where $E_{\tau}$ ($\tau=p,~n$) denotes the quasiparticle energy,
$N_{\tau}$ is the number of particles operator and
\bea
Q_{\tau}=\xi_{\tau}\left[A^{\dag}_{2\mu}(\tau)+
(-)^{\mu}A_{2-\mu}(\tau)\right]~,
\eea
is the quadupole operator written in terms of the normalized two 
quasiparticle excitation operator $A^{\dag}_{2\mu}(\tau)$, 
defined  in a standard way by the angular momentum coupling of two 
quasiparticle creation operators of the same kind.
We describe low lying $2^+$ excitations in terms of the QRPA phonon
operator
\bea
\Gamma^{\dag}_{2\mu}=\sum_{\tau=p,n}\left[
 X_{\tau}A^{\dag}_{2\mu}(\tau)
-Y_{\tau}(-)^{\mu}A_{2-\mu}(\tau)\right]~.
\eea
The QRPA equation of motion
$[H,\Gamma^{\dag}_{2\mu}]=E_{2^+}\Gamma^{\dag}_{2\mu}$
leads to the standard system of equations \cite{Rin80}.
By introducing the following short hand notations
\bea
x_{\tau}=\kappa_{\tau}\xi_{\tau}\xi_{\tau}~,~\tau=p,~n~;~~~
x_{pn}=\kappa_{pn}\xi_{p}\xi_{n}~,
\eea
the QRPA roots can be written as follows
\bea
\label{Epn}
E^2_{2^+}&=&2[E_p^2+E_n^2-E_px_p-E_nx_n
\nn
&\pm&\sqrt{(E_p^2-E_n^2-E_px_p+E_nx_n)^2+
4E_pE_nx_{pn}^2}~]~.
\eea
In the absence of the proton-neutron interaction, i.e. $x_{pn}=0$,
one obtains two roots for neutron and proton systems separately
\bea
\label{energy}
E^2_{2^+}&=&4[E_{\tau}^2-E_{\tau}x_{\tau}]~,~~~\tau=n,~p~.
\eea

\begin{figure}[ht]
\begin{center}
\includegraphics[width=8cm]{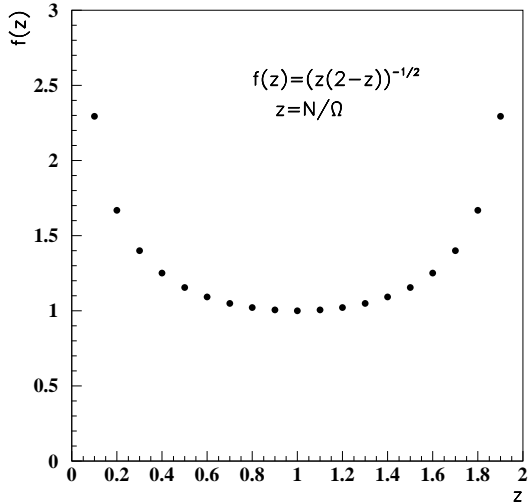}
\vskip-1cm
\caption{
The function defined by Eq. (\ref{fz})}
\label{fig1}
\end{center}
\end{figure}

By using the expression of the quasiparticle energy in terms of the 
Lagrange multipler (we set the single particle energy to zero) and 
pairing gap,
$E_{\tau}=\sqrt{\lambda_{\tau}^2+\Delta_{\tau}^2}$,
one obtains the following simple relation
\bea
\label{ED}
(E_{2^+}/2)^2-\Delta^2_{\tau}&=&\lambda_{\tau}^2-y_{\tau}\Delta_{\tau}~,
\eea
where the coefficient is given by
\bea
\label{yzf}
y_{\tau}=x_{\tau}f(z_{\tau})~.
\eea
Here, we introduced the following universal function
\bea
\label{fz}
f(z_{\tau})&=&\sqrt{1+\left(\lambda_{\tau}/\Delta_{\tau}\right)^2}
=\frac{1}{\sqrt{z_{\tau}(2-z_{\tau})}}
\nn
z_{\tau}&\equiv&N_{\tau}/\Omega_{\tau}~,
\eea
where we used the number of particles condition in terms of the
valence particle number $N_{\tau}$. 
This function is plotted in figure \ref{fig1}. 
One sees that around the middle of the shell ($N_{\tau}=\Omega_{\tau}$)
this function has an almost constant value $f(z)\approx 1$.
Actually for the regions of a shell where $f(z)$ varies a lot 
(that is near the beginning and end) the $2^+$ states are not very 
collective so they are outside our perspective.
Eq. (\ref{energy}) can be rewritten as follows
\bea
\label{theor}
E^2_{2^+}=4[\ov{\Delta}^2_\tau-\ov{\Delta}_{\tau}x_{\tau}]~,~~~
\ov{\Delta}_{\tau}\equiv\Delta_{\tau}f(z_{\tau})~.
\eea

\begin{figure}[ht]
\begin{center}
\includegraphics[width=8cm]{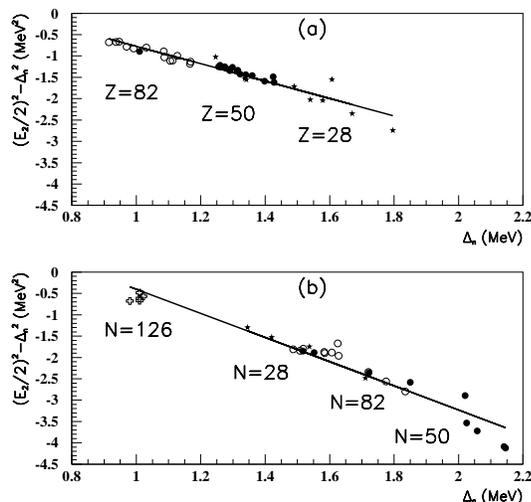}
\vskip-1cm
\caption{
(a) Fit of equation (\ref{ED}) for the neutron-like root
for Z=28 (stars), 50 (dark circles), 82 (open circles) isotopic chains.
(b) Same as in (a), but for N=28 (stars), 50 (dark circles), 
82 (open circles), 126 (open crosses) isotonic chains.}
\label{fig2}
\end{center}
\end{figure}

\begin{figure}[ht]
\begin{center}
\includegraphics[width=8cm]{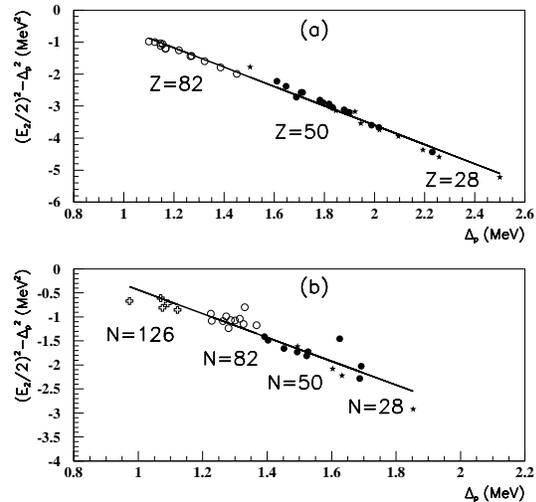}
\vskip-1cm
\caption{Same as in figure \ref{fig2}, but for the proton-like root.}
\label{fig3}
\end{center}
\end{figure}

The fitting procedure of the general relation (\ref{Epn})
showed that the proton-neutron 
interaction $x_{pn}$ has vanishing values along various isotopic and 
isotonic chains, except for the Z=28 isotopic chain, because the
proton-neutron interaction plays an important role for $N\approx Z$ 
nuclei.

Concerning the parameter $y_{\tau}$, we considered two versions. 
In the first approximation we supposed a constant value of 
$y_{\tau}$ in Eq. (\ref{ED}). Later on, we will
consider a mass dependent parameter.

In Ref. \cite{Sch76} the averaged two-body matrix elements
in odd-odd nuclei extracted from experimental data versus the angle 
between the orbital angular momenta of the interacting particles were
analyzed.
As an interesting observation, it turns out that in most cases the shape 
of this dependence is proportional to the function $1/f(z_{\tau})$. 
{\it Thus, in principle the function $f(z_{\tau})$ could be
be compensated by the variation of the parameter $x_{\tau}$, 
keeping in this way the parameter $y_{\tau}$ in Eq. (\ref{yzf})
almost constant.}

In figure \ref{fig2} (a) one sees how the relation (\ref{ED}) is nicely 
fulfilled for the neutron-like root along various isotopic chains
(even for Z=28). It is important to point out that the slope parameter
$y_{\tau}$ has indeed an almost constant value.

\begin{center}
\centerline{Table I}
{Fitting parameters in Eq. (\ref{ED}) with $\tau=n,p$
for isotopic (a) and isotonic (b) chains}

\begin{tabular}{|c|c|c|c|c|}
\hline
Figure & $\tau$ & $\lambda^2_{\tau}$ & $y_{\tau}$ & $\sigma$ \cr
\hline
\ref{fig2} (a) & n & 1.267 & 2.040 & 0.128 \cr
\hline
\ref{fig2} (b) & n & 2.434 & 2.832 & 0.228 \cr
\hline
\ref{fig3} (a) & p & 2.451 & 3.023 & 0.118 \cr
\hline
\ref{fig3} (b) & p & 2.038 & 2.475 & 0.188 \cr
\hline
\end{tabular}
\end{center}

An interesting fact is revealed by figure \ref{fig2} (b),
showing that the same relation is also fulfilled by the isotonic 
chains, i.e., by those neutrons filling the corresponding closed shells.
Thus, the pairing effects of these neutrons are also important
in explaining the low lying $2^+_1$ collective state.

The straight lines in figure \ref{fig2} are defined by the fitting 
parameters $\lambda_{\tau}^2$ and $y_{\tau}$, given in the Table I
for isotopic (a) and isotonic (b) chains.

It is very important to stress on the fact that the same relation is 
also fulfilled for the corresponding proton-like root, as can be seen 
from figure \ref{fig3}. Thus, as in the previous case, for the isotopic  
chains (a) the proton pairing effects play an important role.
It is interesting to point out that the best fit is given in figure 
\ref{fig3} (a), i.e., by protons filling a closed shell.

\begin{figure}[ht]
\begin{center}
\includegraphics[width=8cm]{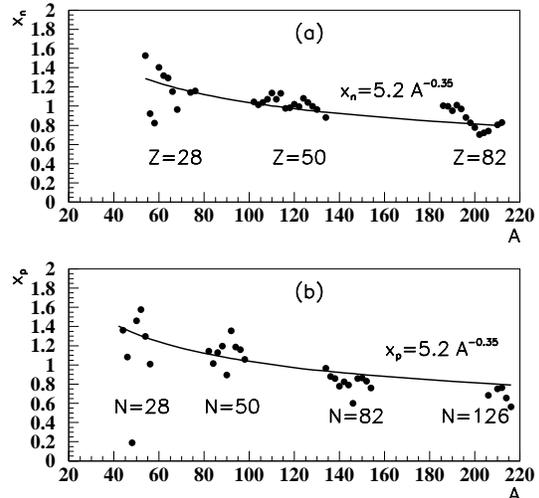}
\vskip-1cm
\caption{
The parameter $x_n$ determined by Eq. (\ref{theor}) with 
$f(z_{\tau})=1$ for isotopic chains (a) and $x_p$  for 
isotonic chains (b).
The curve is given by the common Eq. (\ref{xtau}).}
\label{fig4}
\end{center}
\end{figure}

The next step is to consider a variable parameter $y_{\tau}$,
by using Eq. (\ref{theor}).
In figure \ref{fig4} (a) we plotted by dark circles the parameter $x_n$ 
for various isotopic chains determined by this equation 
{\it by considering $f(z_{\tau})=1$}. In figure \ref{fig4} (b) the 
parameter $x_p$ for various isotonic chains is given by dark circles.
One sees that the fitting curve given by
\bea
\label{xtau}
y_{\tau}=x_{\tau}=5.2~A^{-0.35}~,
\eea
approximates in a satisfactory way the computed values.
We stress the fact that Eq. (\ref{theor}) with $f(z)=1$ is a 
particular case of Eq. (\ref{ED}) with $\lambda_{\tau}=0$, but with
a {\it variable parameter $y_{\tau}$}.
Let us mention here that in Ref. \cite{Uhe66} a much stronger dependence
was derived, but only for the coupling strength 
$\kappa_{\tau}\sim A^{-2.2}$. 

On the other hand, our analysis revealed that the values of the 
parameter $x_p$, but {\it for closed shell isotopic chains}, and 
$x_n$ {\it for closed shell isotonic chains}  have much more 
scattered values than in figure \ref{fig4}.
This means that the "natural" values of the parameter $x_p$ 
{\it along open shell isotonic chains} and $x_n$ {\it along 
open shell isotonic chains} are more  appropriate to be used in 
our systematics.

\begin{figure}[ht]
\begin{center}
\includegraphics[width=8cm]{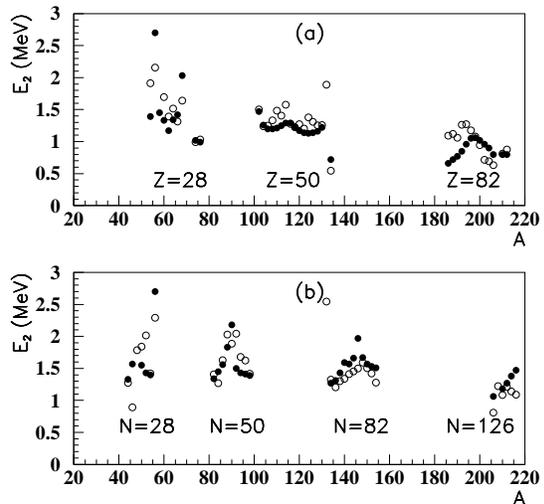}
\vskip-1cm
\caption{
Open circles denote the energies given by Eq. (\ref{theor}) with 
$\tau=n$  for various isotopic chains (a) and with $\tau=p$ for various 
isotonic chains (b). Experimental values are given by dark circles.}
\label{fig5}
\end{center}
\end{figure}

By using the above scalling law (\ref{xtau}) we plotted
by open circles in figure \ref{fig5} (a) the energies computed according 
to  Eq. (\ref{theor}) with $\tau=n$ for various isotopic chains.
In figure \ref{fig5} (b) similar values with $\tau=p$ for
various isotonic chains are given.
One remarks that, except for $Z=28$ isotopes and $N=28$ isotones, the 
experimental values given by dark circles are satisfactorily reproduced.

The fact that both proton and neutron roots are important
is also supported by an analysis of electromagnetic transitions.
The E2 transition matrix element, connecting the excited state $|k\ra$ 
with the ground state $|0\ra$, is given by the following standard  
expression
\bea
\label{T2}
\la 0 || T_2 || k \ra&=&
\sum_{\tau=p,n}e_{\tau}\xi_{\tau}
\left[X_{\tau}^{(k)}+Y_{\tau}^{(k)}\right]
\nn&=&
\sum_{\tau=p,n}e_{\tau}\xi_{\tau}
\sqrt{\frac{2E_{\tau}}{E_{2^+}}}~,
\eea
where $e_{\tau}$ denote the effective proton and neutron charges.
We compared experimental values with the quantity 
$B(E2:k\rightarrow 0)=|\la 0 ||T_2|| k \ra|^2$.

\begin{figure}[ht]
\begin{center}
\vskip-1cm
\includegraphics[width=8cm]{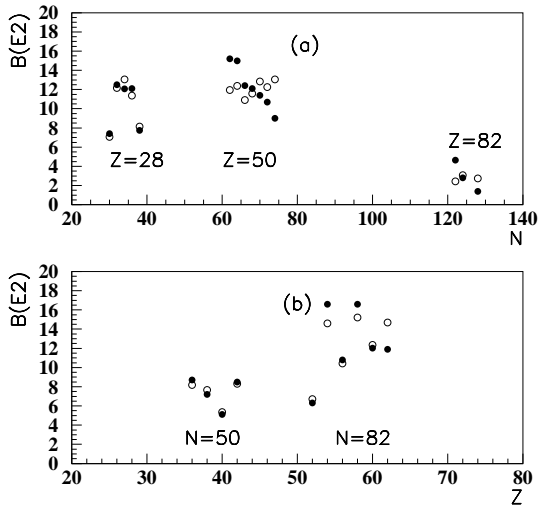}
\caption{(a) B(E2) values versus neutron number for various isotopic 
chains. Experimental values are given by dark circles,
while open circles denote the fit (\ref{T2})
with respect to parameters $a_{\tau}=\sqrt{2}e_{\tau}\xi_{\tau}$.
(b) Same as in (a), but for various isotonic chains.}
\label{fig6}
\end{center}
\end{figure}

In figure \ref{fig6} we plotted B(E2) values for various isotopic
(a) and isotonic chains (b).
By dark circles experimental values are given,
while open circles denote the fit with respect to the parameters 
$a_{\tau}=\sqrt{2}e_{\tau}\xi_{\tau}$.
In Table II we give the fitting parameters $a_{\tau}$ for these chains. 
Indeed, one remarks that both proton and neutron effective
charges have simultaneously nonvanishing and comparable values
with opposite signs, i.e., the pairing effects are very strong for
both kinds of particles.

In conclusion, we have analyzed low lying collective $2^+_1$ states
in semi-magic nuclei, where experimental data concerning energy
levels and B(E2) values are available. We used a toy Hamiltonian
with one proton and one neutron level and a pairing plus quadrupole 
force as the two body interaction. The collective $2^+_1$ state
we described within the spherical QRPA. It turns out that,
except for the Z=28 isotopic chain, the proton-neutron QQ interaction
has vanishing values. 
A very simple relation connecting the $E^+_{2_1}$ energy with the
pairing gap and interaction parameter $x_{\tau}$ 
allows us to describe with a reasonable accuracy experimental data.
This law is fulfilled with a good accuracy for both proton and neutron 
roots. This feature clearly shows the important role played by "inert"
particles inside the closed shells.
This conclusion is also supported by the analysis of B(E2) values.
Indeed, both proton and neutron effective charges with opposite signs
should have nonvanishing values in order to explain the experimental 
data.

\begin{center}
\centerline{Table II}
{Fitting parameters $a_{\tau}=\sqrt{2}e_{\tau}\xi_{\tau}$ 
in Eq. (\ref{T2}) with $\tau=n,p$ for isotopic chains in figure 
\ref{fig6} (a) and isotonic chains in figure \ref{fig6} (b).}

\begin{tabular}{|c|c|c|c|}
\hline
Chain & $a_p$ & $a_n$ & $\sigma$ \cr
\hline
Z=28 &  9.998 &  -6.185 & 0.041 \cr
\hline
Z=50 &  7.036 &  -3.499 & 0.131 \cr
\hline
Z=82 & -2.169 &   2.950 & 0.253 \cr
\hline
N=50 & -4.742 &   7.473 & 0.034 \cr
\hline
N=82 & 81.627 & -84.446 & 0.084 \cr
\hline
\end{tabular}
\end{center}


\centerline{\bf Acknowledgments}

This work was supported by the contracts IDEI-119 and IDEI-180 
of the Romanian Ministry of Education and Research.
We thank prof. R.F. Casten (Yale University),
prof. J. Suhonen (Jyv\"askyl\"a University), and prof. R.J. Liotta 
(KTH, Stockholm) for valuable comments.

\end{document}